\documentclass[pra,twocolumn,english,showpacs,floatfix]{revtex4-1}
\usepackage{graphicx}
\usepackage{amssymb}
\usepackage{color}
\usepackage{amsmath}

\begin{document}
\title{Loading and detecting a three-dimensional Fermi gas in one-dimensional optical superlattice}
\author{Ameneh Sheikhan, Corinna Kollath}

\affiliation{HISKP, University of Bonn, Nussallee 14-16, 53115 Bonn, Germany}

\begin{abstract}
We investigate the procedures of loading and detecting three-dimensional fermionic quantum gases in a one-dimensional optical superlattice potential subjected to a trapping potential. Additionally, we consider the relaxation dynamics after a sudden change of the superlattice potential. We numerically simulate the time-dependent evolution of the continuous system using exact diagonalization of non-interacting fermions. 
During the loading procedure we analyze the occupation of the instantaneous energy levels and compare the situation in a homogeneous system with the trapped one. Strong differences are found in particular in the evolution of excitations which we trace back to the distinct global density distribution.  Starting from an imbalanced state in the superlattice potential, we consider the relaxation dynamics of fermions after a slow change of the superlattice potential and find a bimodule distribution of excitations. To be able to compare with the experimental results we also simulate the measurement sequence of the even and odd local density and find a strong dependence of the outcome on the actual ramp procedure. We suggest how  the loading and detecting procedure can be optimized. 
\end{abstract}

\maketitle

\section{Introduction}
The recent experimental progress to manipulate quantum gases made it possible to study different interesting quantum phenomena. 
The clarification of the mechanism underlying the BCS to Bose-Einstein condensate crossover in different dimensions has been one of the successful advances \cite{KetterleZwierlein2008}. Further, in optical lattices the superfluid to Mott-insulator transition in bosonic gases \cite{BlochZwerger2008} and the liquid to Mott-insulator transition in fermionic gases 
\cite{Schneider2008,Joerdens2008} have been realized. 
However, a lot of interesting quantum phases such as the antiferromagnet, d-wave superconductivity or spin liquids \cite{Lewenstein2007,BlochZwerger2008} have not yet been reached due to the low temperature or more precisely low entropy per particle required \cite{JoerdensTroyer2010, DeLeoParcollet2008}.  For bosons it has been shown that the temperature increases with ramping up an optical lattice  and it can be minimized with tuning the trap frequency \cite{PolletTroyer2008,DolfiTroyer2014}. Thus, the detailed investigation of the loading procedure of a fermionic gas in three-dimension is of utmost importance. 

Moreover, quantum gases have the advantage that they are only very weakly coupled to the environment and can be considered as almost isolated on relatively long time-scales. Thus, quantum gases enable the investigation of the quantum dynamics of isolated systems. A high amount of attention have received so-called quantum quenches. A quantum quench is one specific way to set the system out of equilibrium: starting with a system prepared in an equilibrium state of a chosen initial Hamiltonian, one or several of the system parameters are changed. The state evolves according to the new Hamiltonian. Such a change is either sudden or with a certain ramp speed. The quench velocity of the slow quench plays an important role and has been studied in gapped and gapless phases and across phase transitions relating to the Kibble Zurek mechanism \cite{PolkovnikovSilva2011}. One of the fundamental questions related to a sudden quench is how fast correlations can be formed and whether the isolated system relaxes to a thermal equilibrium state. These questions have been widely studied theoretically both in integrable and non-integrable systems as in Refs.~\cite{Igloi2000, CalabreseCardy2006, Cazalilla2006, Rigol2006, KollathAltman2007,ManmanaMuramatsu2007,BarmettlerRey2008, RigolOlshani2008, Lauchli2008, Cubitt2008, BarmettlerRey2008, BarthelSchollwoeck2008, Rossini2009, Manmana2009, IucciCazalilla2009, MoeckelKehrein2009, Uhrig2009, SotiriadisCardy2010,  Mathey2010, BarmettlerAltman2010, PolkovnikovSilva2011, KollarEckstein2011, Enss2012, Goth2012,  MosselCaux2012, BarmettlerKollath2012,  Natu2013, EsslerKehrein2014, Queisser2014,  AokiWerner2014}.

So far many experiments were conducted for bosonic gases. The observation of the dynamics of bosons in optical lattices \cite{GreinerBloch2002b,Sebbystrabley2007,FoellingBloch2007,BakrGreiner2009,HungChin2010,ShersonKuhr2010,ChenDeMarco2011,RonzheimerBloch2013}
, the prethermalization of one-dimensional condensates \cite{KinoshitaWeiss2006,GringSchmiedmayer2012}, the light-cone-like spreading of correlations after a quench \cite{Cheneau2012} are examples of the experimental activities in this area. Moreover, the dynamics within a superlattice  has been studied in the experiment \cite{TrotzkyBloch2012} and theory \cite{CramerEisert2008, FleschEisert2008, BarmettlerRey2008,BarmettlerAltman2010,Rigol2006}.

Less experiments have been conducted for fermionic quantum gases. Mainly, interaction or lattice quenches \cite{Strohmaier2007,Hackermueller2010,Schneider2012,PertotKollath2014} and spin dynamics have been investigated \cite{KrauserSengstock2012, KoschorreckKoehl2013}. 

In this paper we study the ramp up procedure to load fermionic atoms into first two-dimensional pancakes and then into a one-dimensional superlattice potential. The additional remaining two spatial dimensions are subjected to a harmonic trap. We analyse in detail the loading procedure and suggest improvements for the experiments. Starting from a prepared imbalanced quantum gas, we perform a fast and slow quantum quench and follow the relaxation dynamics of the fermionic gas in the trap. In particular, we find a bimodule density distributions at long times which can be attributed to a situation in which locally the gas is equilibrated, but globally a highly excited state is reached. Additionally, we analyse in detail the detection sequence of the local even-odd density and point out possible improvements to the sequence.

\section{Description of the system}
We study a non-interacting gas of $N=10^5$ (for example $^{40}$K) fermionic atoms with mass $m$ in a three dimensional dipole trap and a one-dimensional optical superlattice along the $x$-direction. The Hamiltonian of the non-interacting atoms at time $t$ is given by 
\begin{eqnarray}
H&=&\int {\it d}{\bf r}\left[\Psi^\dagger({\bf r})\frac{-\hbar^2}{2m}\nabla^2\Psi({\bf r})\right.\nonumber\\
&+&\frac{1}{2}m (\omega_{x}^2(t)x^2+\omega_y^2(t)y^2+\omega_z^2(t)z^2)\Psi^\dagger({\bf r})\Psi({\bf r})\nonumber\\
&+&\left. V(x,t)\Psi^\dagger({\bf r})\Psi({\bf r})\phantom{\int}\right].
\label{eq:Hxyz_time_sec_quant}
\end{eqnarray}
Here $\Psi^\dagger({\bf r})$ and $\Psi({\bf r})$ are the fermionic creation and annihilation operators.
The bichromatic lattice potential along $x$-direction is formed by an infra-red laser beam with wavelength $\lambda_R = \frac{2\pi}{k_R}$ and a green laser beam with wavelength $\lambda_G=\frac{2\pi}{k_G}=\frac{\lambda_R}{2}$. At time $t$ the potential can be described by the superposition of the two lattices 
\begin{eqnarray}
V(x,t)=V_G(t)\cos^2(2k_Rx -\phi(t))-V_R(t) \cos^2(k_Rx),
\label{eq:bi_pot}
\end{eqnarray}
where $V_R(t)$ and $V_G(t)$  are the amplitudes of the red and green lattice potentials at time $t$, respectively. The geometry of the optical superlattice is strongly determined by the phase difference $\phi$ between the two lattices at the position of the atomic cloud. In the experiment the phase can be controlled by detuning the frequency of the green laser from exactly twice the frequency of the red laser. 

 At the position of the atoms the dipole trap can be approximated by a harmonic trap with frequencies $\omega_x$, $\omega_y$ and $\omega_z$. The effective trap frequencies in $y$- and $z$-direction change by varying the potential of the lattice laser beams due to the change of the radial intensity profile. For the red-detuned lattice (red lattice) the frequency of the resulting additional confinement is related to its amplitude by 
\begin{eqnarray}
\omega_{\alpha,R}(t)=\left(\frac{4 V_R(t)}{m \text{w}_{\alpha,R}^2}\right)^{1/2}
\label{eq:omega_red}
\end{eqnarray}
and for the blue-detuned lattice (green lattice) the frequency of the resulting {\it deconfinement} is 
\begin{eqnarray}
\omega_{\alpha,G}(t)=\frac{h}{m\text{w}_{\alpha,G}\lambda_G}\left(\frac{V_G(t)}{E_{r,G}}\right)^{1/4}.
\label{eq:omega_blue}
\end{eqnarray}
Here $\alpha$ denotes the $y$- and $z$-direction and $\text{w}_{\alpha,\gamma}$ is the waist of the corresponding lattice beam. 
The frequency in $x$-direction remains almost constant.

The resulting time dependence of the total effective trap frequencies are 
\begin{eqnarray}
&\omega_{\alpha}(t)=\left[\omega_{\alpha,0}^2+\omega_{\alpha,R}^2(t)-\omega_{\alpha,G}^2(t)\right]^{1/2},\quad\alpha=y,z\nonumber\\
&\omega_x(t)=\omega_{x,0}.
\label{eq:omega_time}
\end{eqnarray}
In order to be able to compare directly to the experiment \cite{PertotKollath2014} we use the initial trapping frequencies in $x$-, $y$- and $z$-direction  $\omega_{x,0}=2\pi\times 38$ Hz, $\omega_{y,0}=2\pi\times 66$ Hz and $\omega_{z,0}=2\pi\times 376$ Hz.
The wavelength of the infra-red laser beam is $\lambda_R = 1064 \mathrm{nm}$ with waists of ${\text w}_{y,R} = 120 \mathrm{\mu m}, {\text w}_{z,R} = 130 \mathrm{\mu m}$ and the wavelength of the green laser beam is $\lambda_G = 532 \mathrm{nm}$ with the waists of ${\text w}_{y,G} = 189 \mathrm{\mu m}, {\text w}_{z,G} = 42 \mathrm{\mu m}$ in $y$- and $z$-direction.

The single particle Hamiltonian (Eq.~\ref{eq:Hxyz_time_sec_quant}) is separable in the different directions ($H=H_x+H_y+H_z$). The simulations are performed using exact diagonalization on the discretized version of the Hamiltonian and the full time evolution of the fermionic gas is determined. If not mentioned otherwise, the space is discretized using $\Delta x=\frac{\lambda_{R}}{50}$ in $x$-direction, $\Delta y=\Delta z=\frac{\lambda_{R}}{6}$ in $y$- and $z$-direction and the time-step is discretized to $\Delta t=0.1$ms. This value is chosen since it corresponds to the steps in which the experimental lattice is regulated in Ref.~\cite{PertotKollath2014}. We discuss the convergence in these parameters later on. 

\section{Loading the atoms into the optical superlattice}\label{sec:loading}
\begin{figure}
\includegraphics[width=.99\columnwidth,clip=true]{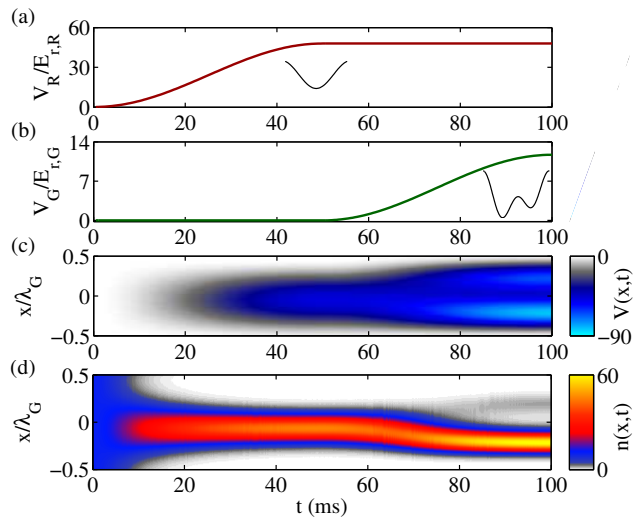}
 \caption{(Color online)
The time dependence of the amplitudes of a) the red and b) the green lattice potential during the loading process. Insets: The black curves are the local potential for $\frac{-\lambda_G}{2}<x<\frac{\lambda_G}{2}$ at a) $t=50 $ms when the red lattice is fully ramped up and b) $t=100$ms when both the red and green lattice are ramped up. c) The evolution of the lattice potential $V(x,t)$ shown for one double well during the loading process. d) The evolution of the local density distribution in one double well during the ramp up process. The density distribution is simulated for the homogeneous gas (case $A$).}
\label{fig:up_hom_pot_dens_mid_20dw}
\end{figure}

In this section we investigate the loading procedure of the non-interacting fermionic gas into the optical superlattice potential along $x$-direction (Fig.~\ref{fig:up_hom_pot_dens_mid_20dw}). We concentrate hereby on loading the atoms first into the red, large wave-length optical lattice. The height of this lattice potential is changed slowly using an $S$-shaped curve of the ramp. Only after the red lattice has reached its full height the short (green) wavelength optical lattice of wavelength $\lambda_G$ is additionally ramped up using the same ramp form. The first part of the ramp therefore also represents the typical loading procedure of Fermi gases into two-dimensional pancakes \cite{PertotKollath2014,GuenterEsslinger2005,FroehlichKoehl2011,Martiyanov2010,SommerZwierlein2012,KoschorreckKoehl2013}. 

The ramp sequence is shown in Fig.~\ref{fig:up_hom_pot_dens_mid_20dw}a,b.
The optical lattice potential in $x$-direction during the ramp process is described by Eq.~\ref{eq:bi_pot}. The intensity $V_\gamma$ at time $t$ follows the $S$-shaped functional form
\begin{eqnarray}
&V_\gamma(t)=V_{\gamma,i}+(V_{\gamma,f}-V_{\gamma,i})\sin^2\left(\frac{\pi(t-t_{\gamma,0})}{2\tau_\gamma}\right),
\label{eq:pot_ramp_time}
\end{eqnarray}
where $\gamma$ can be $R$ and $G$ for the red and green lattice, respectively. $V_{\gamma,i}$ and $V_{\gamma,f}$ are the initial and final optical lattice potentials. The times $t_{\gamma,0}$ and $\tau_\gamma$ are the start time of the ramp and the ramping time for the corresponding lattice and $V_\gamma (t)$ is valid for $t_{\gamma,0}\leq t\leq t_{\gamma,0}+\tau_\gamma$.
 The red lattice is ramped up in $\tau_R=50 $ms from $V_{R,i}=0$ to $V_{R,f}=48 E_{r,R}$ where $E_{r,R}=\frac{\hbar^2 k_R^2}{2m}$ is the recoil energy of the red lattice (see Fig.~\ref{fig:up_hom_pot_dens_mid_20dw}a). Afterwards the green lattice is ramped up for $\tau_G=50 $ms from $V_{G,i}=0$ to $V_{G,f}=11.7 E_{r,G}$, where $E_{r,G}$($=4 E_{r,R}$) is the recoil energy for the green lattice (see Fig.~\ref{fig:up_hom_pot_dens_mid_20dw}b). In the loading process we fix the phase to $\phi=0.12\pi$ such that the final lattice potential consists of a strongly imbalanced superlattice. The evolution of the potential for a single double well is shown in Fig.~\ref{fig:up_hom_pot_dens_mid_20dw}c.

In order to investigate the loading process we perform simulations starting with the ground state of a harmonically trapped gas. We use the initial trapping frequencies in $y$- and $z$-direction as $\omega_{y,0}$ and $\omega_{z,0}$. In order to identify the influence of the trapping potential along the $x$-direction we consider two different cases which we study in details in the two following sections: (A) a homogeneous gas along the $x$-direction, i.e.~$\omega_x=0$ and (B) a trapped gas with a weak trapping frequency $\omega_{x,0}$.

\subsection{Homogeneous gas}
In this section we consider the homogeneous setup with $\omega_x=0$ and open boundary conditions along $x$-direction. Further, we choose the trap frequencies in $y$- and $z$-direction independent of time, i.e.~$\omega_y(t)=\omega_{y,0}=2\pi\times 66 \mathrm{Hz}$ and $\omega_z(t)=\omega_{z,0}=2\pi\times 376 Hz$. Here the space is discretized as $\Delta x=\frac{\lambda_{R}}{200}$. 
We calculate the density distribution for 20 000 particles in 20 double wells (40 single wells of the green lattice) during the ramping up. The evolution of the density distribution of a single double well is shown in Fig.~\ref{fig:up_hom_pot_dens_mid_20dw}d. The initially homogeneously distributed density quickly (time $\approx 10$ ms) localizes at the bottom of the arising lattice wells of the red lattice potential. For the final red lattice depth of $48 E_{r,R}$, the localization of the density is almost perfect and in between the time $40 \mathrm{ms}$ and $50$ ms only slight changes can be identified. When the short wave-length (green) lattice is ramped up starting at $t=50$ms the localized density within one red lattice well splits into two parts. The dominating part of the occupation evolves into the deep odd lattice well of the superlattice, whereas only a very low occupation is transformed to the higher even lattice well.

\begin{figure}
\includegraphics[width=.99\columnwidth,clip=true]{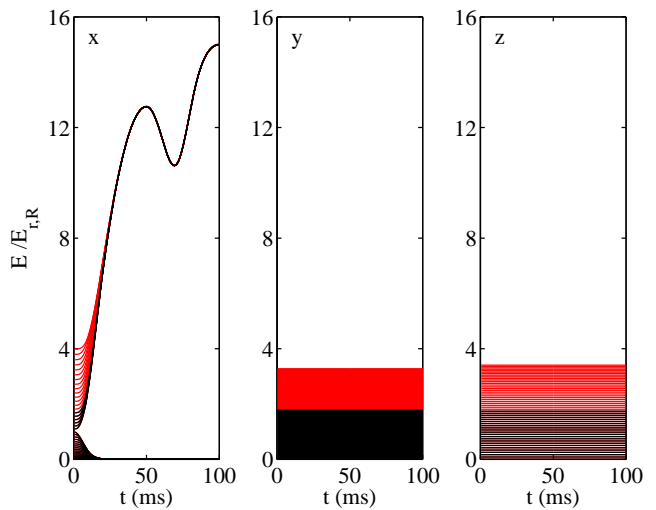}
 \caption{(Color online) Time evolution of the instantaneous eigenenergies in $x$-, $y$- and $z$-direction during the ramping up process for the homogeneous case of 20 double wells with open boundary condition. The energy levels are shifted to have a minimum energy equal to zero at any time. The black lines are occupied levels and the red (grey) lines are unoccupied levels.}
\label{fig:up_hom_el_xyz_20dw_yzcte}
\end{figure}

\begin{figure}
\includegraphics[width=.99\columnwidth,clip=true]{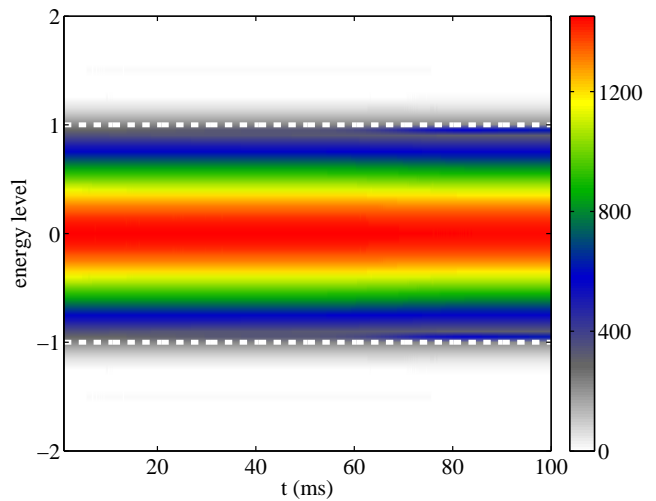}
 \caption{(Color online) Time evolution of the occupation of the eigenenergy levels of $H_x$ integrated over the $y$- and $z$-direction for the case of the  homogeneous system with 20 double wells and 20000 particles during the ramp up process. The vertical axis is the quasi-momentum index of the eigenenergy levels in $x$-direction and the first Brillouin zone of the red lattice lies between -1 and 1.}
\label{fig:up_hom_px_20dw}
\end{figure}
 
In order to monitor the excitations which are created during this ramping process, we determine the time evolution of the occupation of the instantaneous energy eigenlevels (the eigenenergies at each time $t$). Since the Hamiltonian does not change in time along the $y$- and $z$-direction and is separable, the eigenvalues of the parts $H_y$ and $H_z$ and their occupation remain constant over the entire ramp as can be seen in Fig.~\ref{fig:up_hom_el_xyz_20dw_yzcte}y,z. In contrast along the $x$-direction the ramp up of the lattice changes strongly the energy levels (Fig.~\ref{fig:up_hom_el_xyz_20dw_yzcte}x).
During the turning on of the red lattice Bloch bands develop. Each band contains $20$ energy levels (due to the choice of finite length of 20 red lattice wells). The bands are well separated already after approximately $10$ms. A drastic narrowing of the bands is observed. The switching on of the green lattice shifts the position of the bands but almost no mixing of the two Bloch bands is seen. In $x$-direction the evolution of the occupation of the energy levels depends strongly on the ramp procedure. For the chosen one, the evolution is shown in Fig.~\ref{fig:up_hom_el_xyz_20dw_yzcte}x, where occupied levels are represented as black lines and in Fig.~\ref{fig:up_hom_px_20dw}. The initial distribution due to Pauli blocking occupies around 26 lowest energy levels in $x$-direction and almost no change can be seen during the ramp. This means that the ramp procedure is almost adiabatic with respect to the $x$-direction. Comparing the final occupied energies in the different directions, one sees that even though the dynamics is almost adiabatic in $x$-direction, a highly excited final state is created. The ground state of the configuration at the end of the ramp would have lower occupation in $x$-direction and more levels occupied in $y$- and $z$-direction. This is caused by the decoupling of the different lattice directions. In an experiment a finite interaction or misalignment of lattices would cause a distribution of the fermionic occupations between the different lattice directions. 
Let us note that we need to be careful with the interpretation of our simulations at the initial times, since finite size and discretization effects might play an important role. We will study a large system subjected to a trap in the next section. 

\subsection{Trapped gas}

\begin{figure}
\includegraphics[width=.99\columnwidth,clip=true]{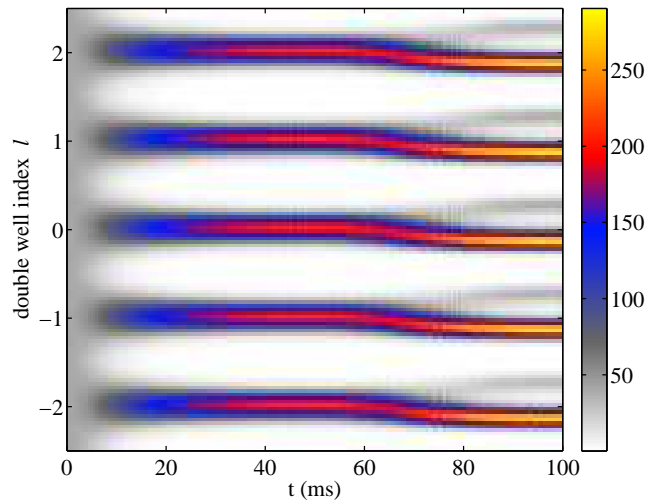}
 \caption{(Color online) Time evolution of the local density for $N=10^5$ fermions with $250$ double wells during the loading process (Case B). For better visibility only 5 double wells in the center of the trap are shown.}
\label{fig:up_trap_dens_250dw_5}
\end{figure}

\begin{figure}
\includegraphics[width=.99\columnwidth,clip=true]{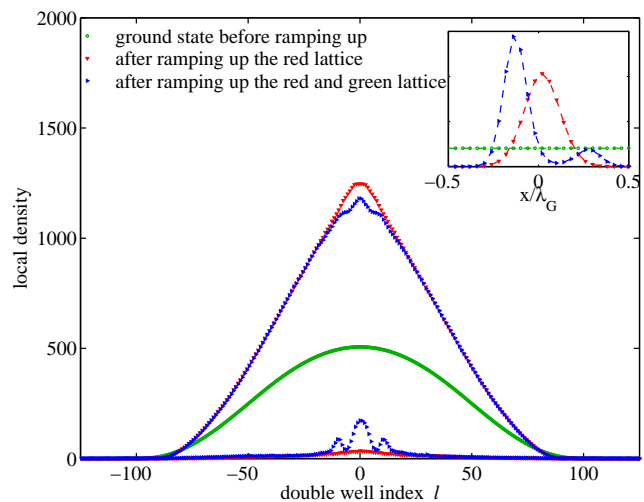}
 \caption{(Color online) The local density of the wells integrated over $y$- and $z$-direction and over the wells at different times of the ramping process for $10^5$ fermions in the harmonic trap (case B) with $250$ double wells. The ground state before and after the loading process are also shown. For each double well index there are two points showing the number of particles in the even and odd well. The odd wells are deeper and thus, have higher density. The inset shows the local density inside one double well in the center of the trap.}
\label{fig:up_trap_dens_250dw_diff_t}
\end{figure}

In the presence of an additional trapping potential along the $x$-direction, the dynamics of the Fermi gas changes.
We take the full time-dependence of the different trap frequencies given by Eq.~\ref{eq:omega_time} into account.
Here the time step used in the simulations is $\Delta t=0.5$ms which is still much smaller than typical time scales of the induced changes beside at the initial time. Fig.~\ref{fig:up_trap_dens_250dw_5} shows a zoom of the evolution of the density distribution during the ramping up. The density distribution starts from a Thomas-Fermi like profile at the initial time $t=0$ (cf.~green curve in Fig.~\ref{fig:up_trap_dens_250dw_diff_t}). This continuous distribution is split by increasing the red lattice potential and after an initial ramp up ($t\leq 20$ms) the Fermions mainly occupy the minima of the red lattice wells. The actual density in a certain lattice well depends on the position of the lattice well in the trap (cf.~$t=50$ms line in Fig.~\ref{fig:up_trap_dens_250dw_5} and the red $\triangledown$ in Fig.~\ref{fig:up_trap_dens_250dw_diff_t}). In particular, the central wells contain more particles than the boundary wells, in which the density falls off to zero. The subsequent splitting of these red lattice wells into the superlattice configuration is almost adiabatic with respect to the local splitting. However, in the central wells, due to the initially higher density, the dynamics induces occupation in the even wells of the green lattice (cf.~$t=100$ms in Fig.~\ref{fig:up_trap_dens_250dw_5} and the blue $\triangleright$ in Fig.~\ref{fig:up_trap_dens_250dw_diff_t}).

\begin{figure}
\includegraphics[width=.99\columnwidth,clip=true]{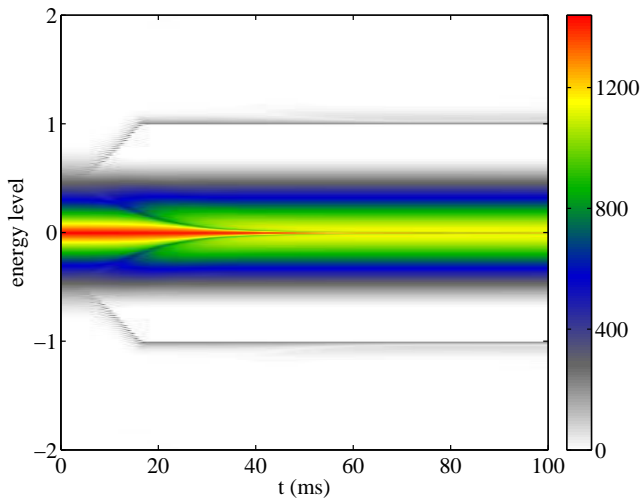}
 \caption{(Color online) Evolution of the energy level occupation in $x$-direction integrated over $y$- and $z$-direction during the ramping up process for $N=10^5$ fermions in the harmonic trap (case B) with $250$ double wells. The vertical axis is the quasi-momentum index of the eigenenergy levels in $x$-direction.  Once the bands are separated, the first Brillouin zone of the red lattice lies between -1 and 1.}
\label{fig:up_trap_px_250dw}
\end{figure}

\begin{figure}
\includegraphics[width=.99\columnwidth,clip=true]{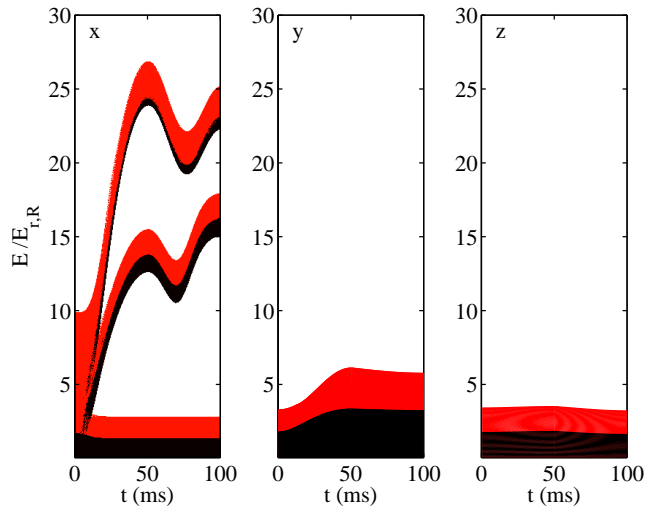}
 \caption{(Color online) Time evolution of the instantaneous eigenenergies in $x$-, $y$- and $z$-direction during the ramp up process for the harmonic trap (case B) with $250$ double wells. The energy levels are shifted to have minimum energy equal to zero at any time. The black lines are occupied levels and the red (grey) lines are unoccupied levels for $N=10^5$ fermions.}
\label{fig:up_trap_el_xyz_250dw}
\end{figure}

\begin{figure}
\includegraphics[width=.99\columnwidth,clip=true]{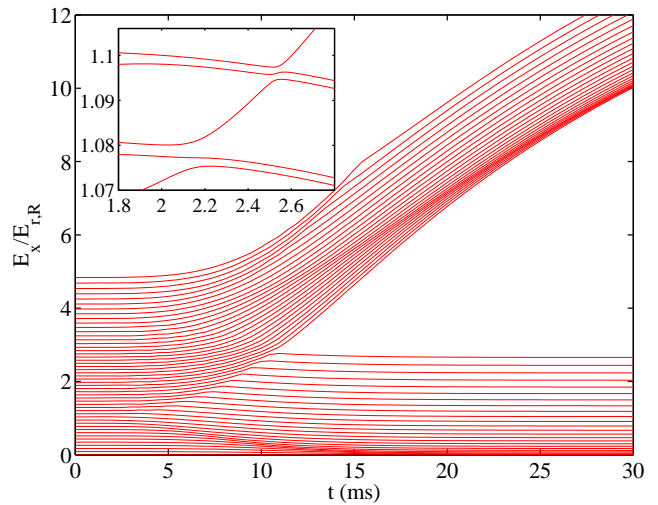}
 \caption{(Color online) Zoom of the time evolution of the instantaneous eigenenergies in $x$-direction during the ramping up process for the harmonic trap (case B) with $250$ double wells. The energy levels are shifted to have minimum energy equal to zero at any time. Only 1 out of 10 energy levels is plotted. One can see the level crossings at the beginning of the ramp process. The inset shows the avoided level crossing of five neighbouring energy levels.}
\label{fig:up_trap_el_x_250dw_10}
\end{figure}

In the following we connect the dynamics to the change of the occupation of the instantaneous energy levels shown in Figs.~\ref{fig:up_trap_px_250dw} and \ref{fig:up_trap_el_xyz_250dw}. As in the case of the homogeneous gas, the energy levels in $x$-direction respond strongly to the ramp up of the red lattice. Due to the time-dependent change of the trap frequencies in $y$- and $z$-direction, also the energy levels of $H_y$ and $H_z$ change. The change in $y$- and $z$-direction is found to be very smooth and the level occupation shows an adiabatic following of the levels in these directions. 
In contrast, along the $x$-direction Bloch bands arise around approximately 10 ms, which separate clearly around $20 $ms. In contrast to the homogeneous case, we see that the energy levels show a strong mixing which is due to the trap along $x$-direction. In particular, in Fig.~\ref{fig:up_trap_el_xyz_250dw} we see that occupied (black) levels arise from the lowest bands and split into Bloch bands around $10 $ms (Here we show three of the Bloch bands).  If the ramping up was adiabatic, then we would expect to have all final states in the lower levels. However, there are many avoided level crossings (Fig.~\ref{fig:up_trap_el_x_250dw_10}) in the beginning (before $10 $ms) and the system is excited to higher levels (Fig.~\ref{fig:up_trap_el_xyz_250dw}). This can be seen well in the occupation of the instantaneous energy levels in Fig.~\ref{fig:up_trap_px_250dw}. An adiabatic ramping would mean no change in the occupation. However,
 we see that a strong change of the distribution is seen up to approximately times of the order of $40$ms. Thus, we see that in the trapped system the dynamics is strongly non-adiabatic caused by the non-locality of the eigenstates. The ramp up scale is too fast in order to assure a redistribution of the entire atomic cloud.

\section{Relaxation dynamics during and after a quench}
In this section we study the evolution of the prepared imbalanced gas after a sudden or slow ramp down of the red lattice.
This change often called a quantum quench, sets the system out of equilibrium and induces a dynamics. 

\begin{figure}
\includegraphics[width=.99\columnwidth,clip=true]{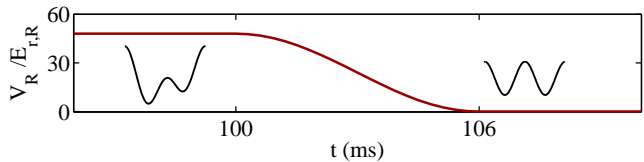}
 \caption{(Color online) The time dependence of the red lattice potential during the slow quench ($0<t<6 $ms). The black curves inside are the local potential for $\frac{-\lambda_G}{2}<x<\frac{\lambda_G}{2}$ before and after the quench. For sudden quench the ramp shape is the same but with smaller ramp time $\tau= 0.5$ms.}
\label{fig:down_gr}
\end{figure}

We start with the initial patterned density distribution with alternating empty and occupied wells in the bichromatic superlattice generated by the loading process described in section \ref{sec:loading}. By switching off the red lattice potential, the double well lattice structure turns into the normal green optical lattice. 
We assume the amplitude of the red lattice ramps down to zero with the same form as given in Eq.~\ref{eq:pot_ramp_time} (cf.~Fig.~\ref{fig:down_gr}). Two different ramp times will be considered (i) $\tau_r=0.5 $ms for an almost abrupt quench which is faster than the tunneling times of the double wells and (ii) $\tau_r=6 $ms for a slow quench which is much larger than the final tunneling times in the green lattice. Some properties of the subsequent relaxation dynamics following these quenches have been analysed by us and collaborators in Ref.~\cite{PertotKollath2014}. We complement these results here by a more detailed analysis showing additional results.  

\begin{figure}
\includegraphics[width=.99\columnwidth,clip=true]{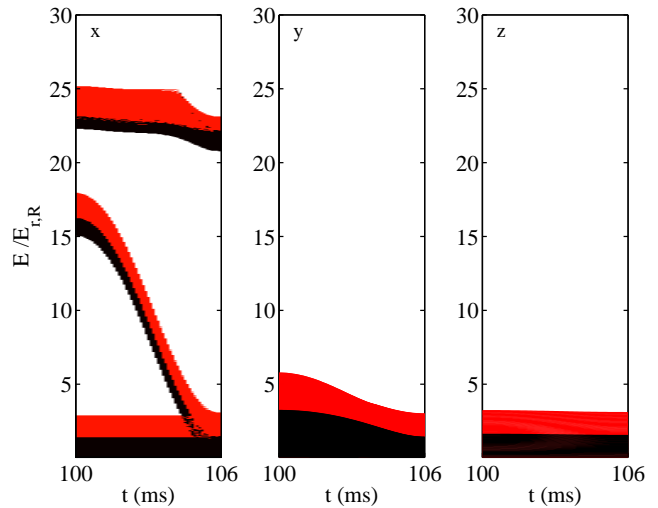}
 \caption{(Color online) Time evolution of the instantaneous eigenenergies in $x$-, $y$- and $z$-direction during the slow quench for the harmonic trap with $250$ double wells. The energy levels are shifted to have minimum energy equal to zero at any time. The black lines are occupied levels and the red (grey) lines are unoccupied levels for $N=10^5$ particles. }
\label{fig:down_trap_el_xyz_250dw}
\end{figure}

In Fig.~\ref{fig:down_trap_el_xyz_250dw} the time evolution of the instantaneous eigenenergies in $x$-, $y$- and $z$-direction are shown. The black lines are the occupied levels during the slow quench. In $x$-direction the lowest two bands join at the end of the ramping down of the red lattice potential. The higher bands remain well separated from the two lower bands since they correspond to higher Bloch bands of the green lattice potential which is still present. During the last stage of the ramp down, many level crossings occur in between the lowest two bands. 

\begin{figure}
\includegraphics[width=.99\columnwidth,clip=true]{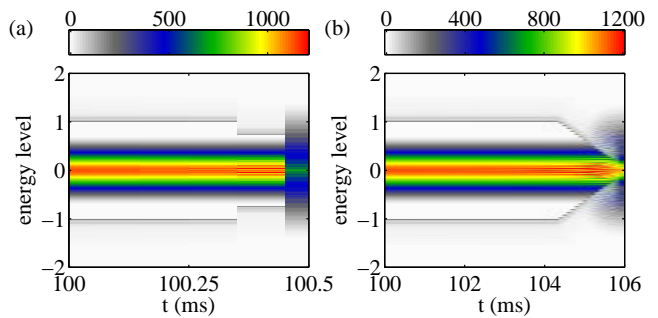}
 \caption{(Color online) Time evolution of the energy level occupation in $x$-direction integrated over $y$- and $z$-direction during (a) abrupt quench and (b) slow quench. $N=10^5$ fermions in the harmonic trap with $250$ double wells. The scale of the vertical axis is chosen as in Fig.~\ref{fig:up_trap_px_250dw}.}
\label{fig:down_trap_band}
\end{figure}

\begin{figure}
\includegraphics[width=.99\columnwidth,clip=true]{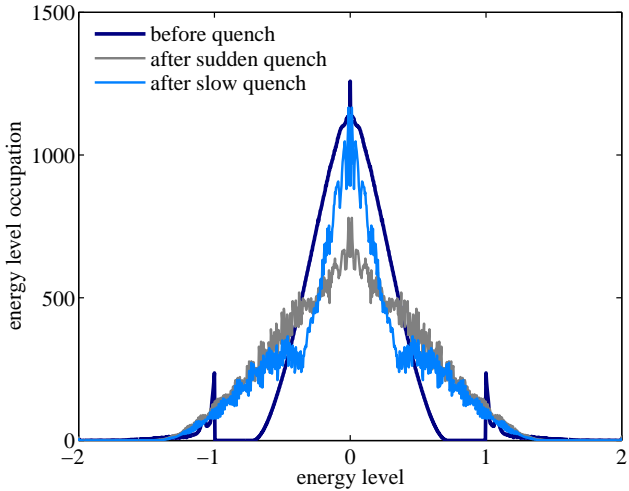}
 \caption{(Color online) The energy level occupation in $x$-direction integrated over $y$- and $z$-direction before and after the abrupt and slow quench. The scale of the horizontal axis is chosen as in Fig.~\ref{fig:up_trap_px_250dw}.}
\label{fig:steady}
\end{figure}

In Fig.~\ref{fig:down_trap_band}a and b the time evolution of energy level occupation is plotted for the almost abrupt and slow quench, respectively. In both cases the occupation remains almost constant over a large percentage of the quench time. Only during the last third of the quench time changes in the occupation are evident. This is due to the initially large gap in between the Bloch bands which prohibits tunneling. Only when the lowest two Bloch bands join, the occupation changes and a redistribution occurs. The final distribution of the occupation after the different quenches is shown in Fig.~\ref{fig:steady}. In the case of the almost abrupt quench many excitations are created. This is reflected in the broad distribution created up to relatively large levels. In comparison during the slow quench, more fermions are transferred to the low energy levels. This process is signalled by the clear peak in the energy level occupation around zero. However, surprisingly the distribution shows a double structure and the number of excitations of the high energy levels ($>0.5$) are comparable to the number of excitations in this region generated by the almost sudden quench. 

In the following we study the relaxation dynamics of the system at different time scales after the almost abrupt and slow quench starting from these highly excited states. 
 
\subsection{One-dimensional tight binding model}

Before presenting numerical results for the dynamics after the quench in three dimensions we start with a toy model for fermions in one dimension. The Hamiltonian is a tight-binding model with tunneling amplitude $J$ and energy offset $\Delta$ between neighbouring sites 
\begin{eqnarray}
H=\sum_{j=1}^L[-J c_j^\dagger c_{j+1} -J c_{j+1}^\dagger c_j  + (-1)^j \frac{\Delta}{2}c_j^\dagger c_j]
\end{eqnarray}
where $L$ is the number of sites. This model has previously been investigated \cite{TrotzkyBloch2012, CramerEisert2008, FleschEisert2008, BarmettlerRey2008, Rigol2006} for bosonic atoms. For completeness we summarize some of the results in the following. For $\Delta=0$, the energy band is the cosine-shaped band. At finite $\Delta$ a gap opens and splits the energy band. In the limit of $\frac{\Delta}{J}\to \infty$ and half filling ($N=\frac{L}{2}$) only odd sites are occupied by one fermion each. For a smaller energy offset, this imbalance decreases until for the homogeneous lattice without energy offset, a homogeneous density distribution is reached. 

The system can be excited by a sudden quench of the energy offset from a finite value $\Delta$ to zero. The subsequent time evolution of the local density $n_j=\langle c_j^\dagger c_j\rangle$ is determined by
\begin{eqnarray}
n_j(t)=\frac{1}{2}-(-1)^j\frac{1}{L}\sum_{k=1}^{L/2}\tanh\theta_k \cos\left(\frac{4Jt}{\hbar}\cos\frac{2\pi k}{L}\right)
\label{eq:n_j}
\end{eqnarray}
where $\sinh\theta_k=\frac{\Delta}{4J|\cos\frac{2\pi k}{L}|}$. Thus, this dynamics depends on the initial energy offset by the function $\tanh\theta_k$. 

We define the averaged odd-even contrast as $C_{oe}=\frac{N_o-N_e}{N}$ where $N_o$ and $N_e$ are the total particle numbers in odd sites and even sites, respectively. 
This quantity can be measured in the experiment \cite{Sebbystrabley2007,TrotzkyBloch2012, PertotKollath2014,FoellingBloch2007,GreifEsslinger2013} (section \ref{sec:exp_detect}). In the limit of a large initial energy offset ($\frac{\Delta}{J}\to\infty$) and infinite size lattice, the averaged odd-even contrast is $C_{oe}(t)=J_0(\frac{4Jt}{\hbar})$ where $J_0$ is the zeroth-order Bessel function of the first kind (same as what found for bosons \cite{CramerEisert2008, FleschEisert2008}). The averaged contrast is maximum in the beginning $C_{oe}(0)=1$, then relaxes symmetrically about $C_{oe}=0$ with the period of $T=\frac{2\pi \hbar}{4J}$ and decays asymptotically in time as $t^{1/2}$. 
If $\frac{\Delta}{J}$ is finite the initial odd-even contrast is less than $1$ and it follows similarly damped oscillations than the Bessel function in time. It differs mainly in a lower amplitude and the details of the damping.

Note that this tight-binding model only covers the potential imbalance and not the exact form of the double well which might include a dimerized hopping. Further, higher Bloch bands are completely neglected and a pure one-dimensional system without additional trapping potential is assumed. Nevertheless, we will recover some of the features in the full relaxation dynamics.

\subsection{Short time dynamics: Relaxation of the local imbalance}
\label{sec:shorttime}
In this section we determine the time evolution of the density distribution of the full model. By integrating the density distribution over $y$- and $z$-direction and also inside each well $i$ we calculate the number of particles $n_i$ in the two dimensional sheet corresponding to the well $i$. Similar to the tight-binding model we probe the density imbalance of the odd-even wells $\Delta n_l=n_{2l+1}-n_{2l}$ (where $l$ is the double well index) and the averaged odd-even contrast $C_{oe}$ which is averaged over the trap.
 
\begin{figure}
\includegraphics[width=.99\columnwidth,clip=true]{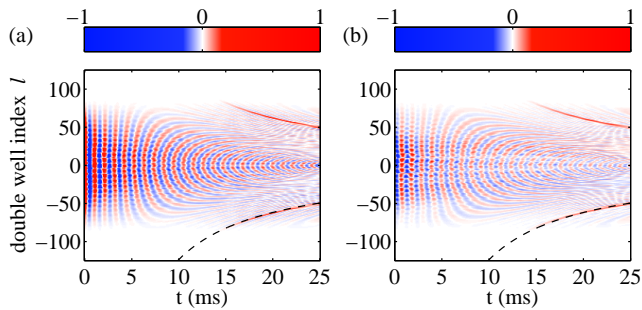}
 \caption{(Color online) Evolution of the local odd-even density imbalance rescaled by its maximum value after (a) the almost abrupt quench and (b) the slow quench for $N=10^5$ fermions in the harmonic trap with $250$ double wells. The dashed lines show the predicted Bloch oscillations.}
\label{fig:dyn_diff_sl_su}
\end{figure}

\begin{figure}
\includegraphics[width=.99\columnwidth,clip=true]{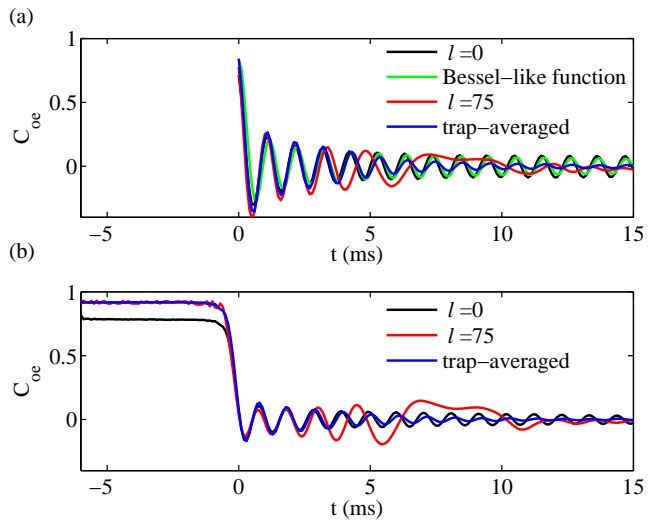}
 \caption{(Color online) a) Evolution of the local even-odd contrast $\left(c_{l}=\frac{n_{2l+1}-n_{2l}}{n_{2l+1}+n_{2l}}\right)$ in the center of the trap ($l=0)$, the boundary of the trap ($l=75$), and the trap averaged contrast $C_{oe}$ after the almost abrupt quench. The contrast in the center fits well with the Bessel-like function predicted from the tight-binding calculation for the homogeneous system. b) Evolution of the trap averaged contrast $C_{oe}$ during and after the slow quench. For both cases the quench ends at $t=0$.}
\label{fig:dyn_cont_sl_su}
\end{figure}

Fig.~\ref{fig:dyn_diff_sl_su}a shows the evolution of the odd-even density imbalance $\Delta n_l$ rescaled by $n_0(t=0)$ after the almost abrupt quench. The central well closely follows the expected Bessel-like function of the one-dimensional tight-binding model with $\frac{\Delta}{J}=4$ in Eq.~\ref{eq:n_j} (Fig.~\ref{fig:dyn_cont_sl_su}a). The period in the Bessel function is $T=\frac{2\pi \hbar}{4J_G}$ where $J_G$ is the tunneling amplitude of the green lattice. We determine the tunneling amplitude from a band structure calculation ($J_G=1.3085\times 10^{-2} E_{r,G}$). The resulting period is $T\approx 1.08 $ms which has slight deviation of $5\%$ from the period obtained in the numerical results of the full model. This slight discrepancy is due to the chosen spatial discretization of $\Delta x=\frac{\lambda_R}{50}$ and becomes smaller for smaller discretizations.
The oscillations of the local odd-even imbalance at the boundaries of the cloud (at $l \approx \pm 70$) show a considerable slowing down and Bloch oscillations appear. The spreading of the period of the oscillations leads in the trap-averaged odd-even contrast to a stronger damping in time (the blue curve in Fig.~\ref{fig:dyn_cont_sl_su}a). Both the slowing down of the tunneling and the onset of Bloch oscillations are due to the local energy offset between neighbouring wells caused by the trapping potential. The position of the maximum contrast closely follows the local Bloch period given by $T_B(l)=\frac{8h}{m \omega_x^2\lambda_R^2l}$ using the local offsets caused by the trap. The slight oscillating features along the $x$-direction are due to the initial non-adiabatic preparation of the atoms in the superlattice. As can be seen in Fig.~\ref{fig:up_trap_dens_250dw_diff_t}, the density distribution after the quench has a low occupation in the upper Bloch band, which shows strong oscillatory structures along $x$-direction.

In Fig.~\ref{fig:dyn_diff_sl_su}b the time evolution of the odd-even density imbalance $\Delta n_l$ rescaled by $n_0(t=0)$ after the slow quench is shown. The general features of the relaxation dynamics after the slow quench are very similar to the almost abrupt quench at short times. In particular, a damped oscillatory behaviour is observed. The main difference is the initially lower amplitude. This can be better seen in Fig.~\ref{fig:dyn_cont_sl_su}b, where the time evolution of trap-averaged odd-even contrast during and after the slow ramp down is plotted. The odd-even contrast at the beginning of the ramp ($t=-6$ms) is $0.92$ which is the contrast after initial non-adiabatic preparation. For the almost abrupt quench the value at time $t=0$ms is very close to this value. In contrast, during the slow quench, the odd-even contrast is much lower and the following oscillations have a much smaller amplitude which is quickly damped out. The different behaviour is related to the more efficient transfer of the fermions in the last stage of the quench to low energy levels. These are mainly the energy levels which correspond to the local redistribution of the fermions, whereas higher energy levels which are still excited almost as intensively as in the abrupt quench correspond to extended states. Thus, the width of the density distribution has not yet varied considerably during the slow quench, such that no global relaxation has taken place. This motivates us to investigate in more detail the global relaxation taking place at longer times in the next section.

\subsection{Breathing modes}
\begin{figure}
\includegraphics[width=.99\columnwidth,clip=true]{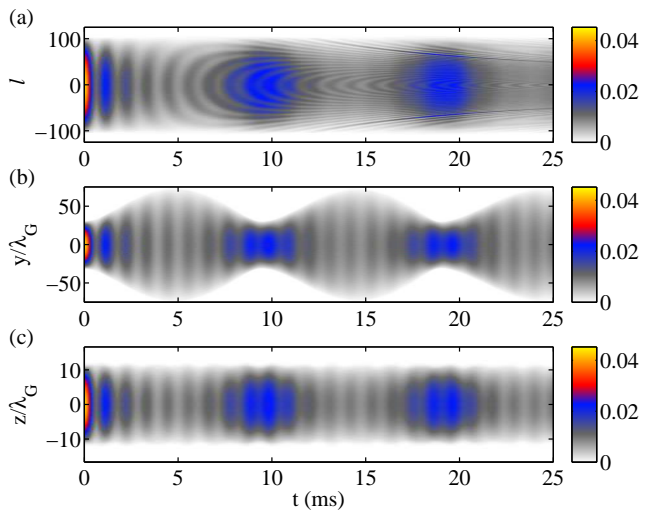}
 \caption{(Color online) Dynamics of the local density in $y$-direction ($x=z=0$) for the discrete tight-binding model (see Appendix \ref{app:tightbinding}) after the abrupt quench for $10^5$ particles in the trap with $500$ sites.}
\label{fig:breathing}
\end{figure}

Additionally to the dynamics along the $x$-direction, the change of the effective trapping frequency in $y$- and $z$-direction (Eqs.~\ref{eq:omega_time}) due to the increase of the lattice along $x$-direction induces oscillations of the density distribution. In Fig.~\ref{fig:breathing} b) we show the local density along $y$-direction in the center of the cloud ($x=z=0$) at different times. We perform the simulation for a discrete tight-binding model as explained in the appendix \ref{app:tightbinding}. A breathing of the density distribution with a breathing period $T\approx 10 $ms (corresponds to the breathing frequency $2\omega_y\approx 2\pi\times96$ HZ) can be clearly seen. This strong change in the density distribution is also reflected in the evolution of the distributions along the other directions as seen in Fig.~\ref{fig:breathing} a and c. Let us note, that this could not be seen in the previously discussed observables as the local density imbalance, since these were integrated over the $y$- and $z$-direction.

\subsection{Long time: Global relaxation dynamics}

In this section we focus on the {\it global} relaxation dynamics of the cloud after the quench. This is motivated by our finding in the previous section that the local odd-even density imbalance relaxes algebraically fast to zero, a possible thermal equilibrium value. The initial density profile along $x-$direction before the quench is relatively broad since only every second lattice well is occupied. The ground state profile in the lattice after the quench would be much more narrow. In the presence of the trap the global dynamics after the quench acts to redistribute the density profile. 
A good candidate to probe such a global relaxation inside the trap is the quasi-momentum distribution of the corresponding released system measured in experiments by the band-mapping technique. In the simulations we obtain the quasi-momentum distribution by a direct projection onto the homogeneous quasi-momenta. This distribution would not evolve with time in the absence of a trapping potential and is, thus, very well suited to measure the induced global relaxation.  

\begin{figure}
\includegraphics[width=.99\columnwidth,clip=true]{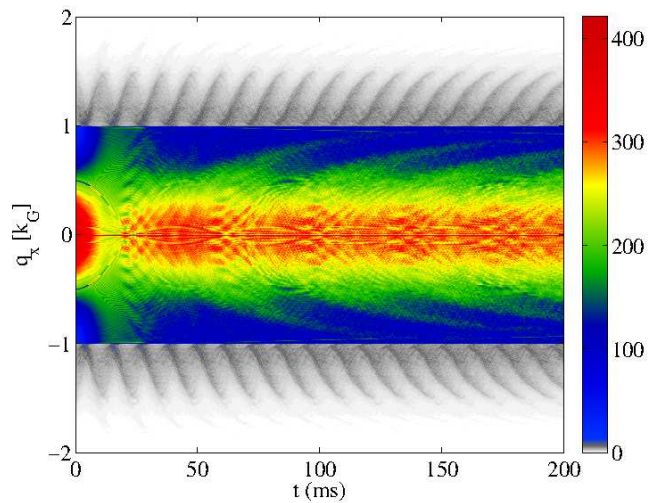}
 \caption{(Color online) Evolution of the homogeneous quasi-momentum distribution in $x$-direction integrated over $y$- and $z$-direction after the slow quench. It is simulated for $10^5$ particles in the trap with $250$ double wells. The scale of the vertical axis is chosen as in Fig.~\ref{fig:up_trap_px_250dw}.}
\label{fig:dyn_band_sl}
\end{figure}

\begin{figure}
\includegraphics[width=.99\columnwidth,clip=true]{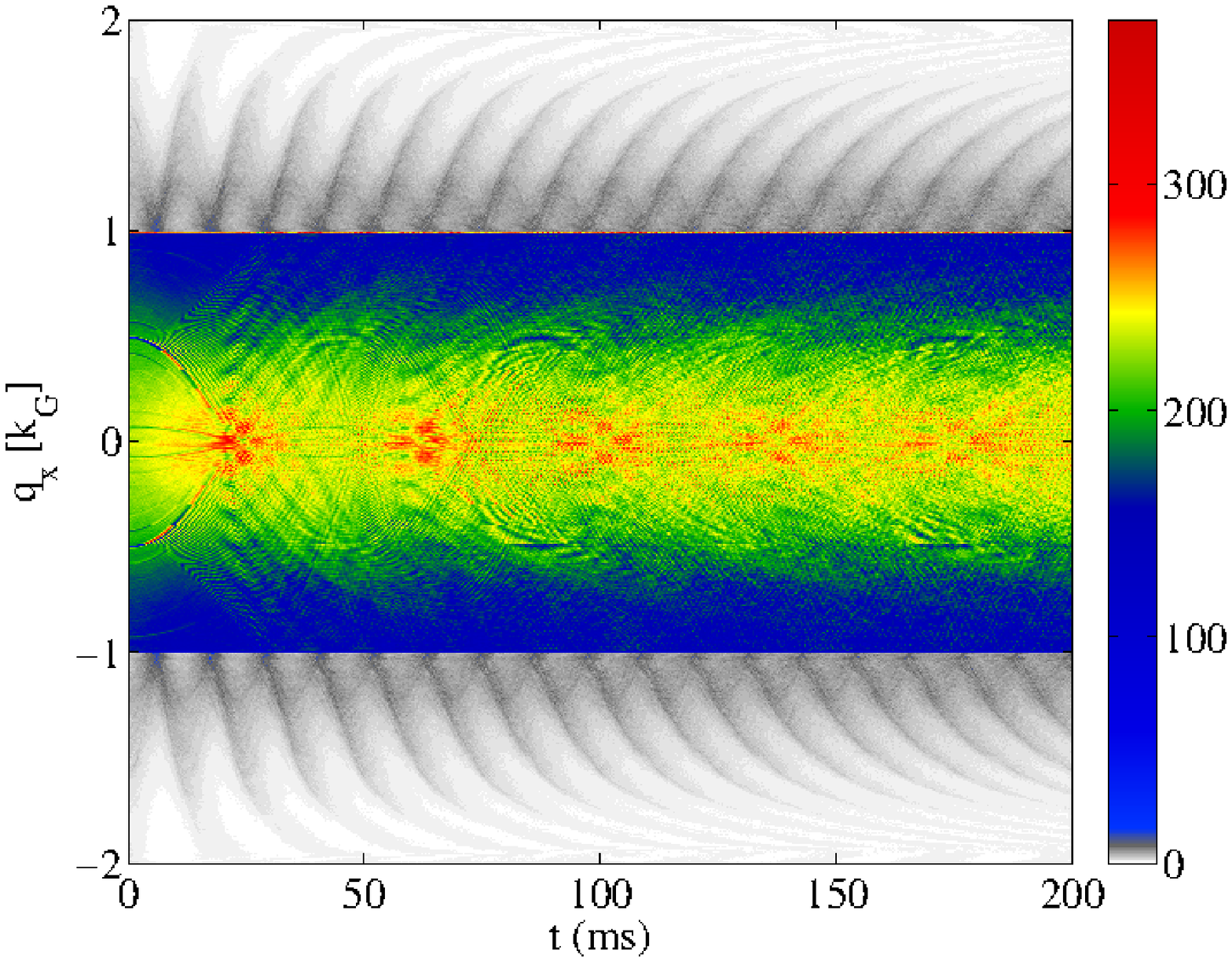}
 \caption{(Color online) Evolution of the homogeneous quasi-momentum distribution in $x$-direction integrated over $y$- and $z$-direction after the almost abrupt quench. It is simulated for $10^5$ particles in the trap with $250$ double wells. The scale of the vertical axis is chosen as in Fig.~\ref{fig:up_trap_px_250dw}.}
\label{fig:dyn_band_su}
\end{figure}

In Fig.~\ref{fig:dyn_band_sl}, the time evolution of the homogeneous quasi-momentum distribution is shown after the slow quench. 
An initial focusing of the occupation towards the center ($t\leq 10$ ms) is observed for the quasi-particles in the first Brillouin zone ($|q|<k_G$). Subsequently, they are accelerated to higher momenta which is signalled by the outgoing diagonals from the center. The oscillations between low and high momenta which leads to the repetition of such diagonals, make the evolution resemble a fish-bone-like structure.
A small fraction of particles are in the higher bands ($\approx 10\%$). In the second Brillouin zone ($k_G<|q|<2 k_G$) a similar acceleration is seen which takes place with a different frequency. In both Brillouin zones, these oscillations seem to persist over a long time (several hundreds of ms). For the case of the almost abrupt quench the main feature resemble those of the slow quench. However, they are much less pronounced (Fig.~\ref{fig:dyn_band_su}) and faster oscillating structures occur.

In order to quantify the dynamics better we define the contrast within the first and second Brillouin zones as 
$C_1=\frac{N_1^1-N_1^2}{N_1^1+N_1^2}$ and $C_2=\frac{N_2^1-N_2^2}{N_2^1+N_2^2}$ where $N_j^1$ is the number of atoms in the first half of the $j-$th Brillouin zone i.e.~the occupation of the quasi-momentum states in the interval $\frac{(2j-2)\pi}{\lambda_G}<|q|<\frac{(2j-1)\pi}{\lambda_G}$ and $N_j^2$ is the number of atoms in the second half of the $j-$th Brillouin zone i.e.~the occupation of the quasi-momentum states in the interval $\frac{(2j-1)\pi}{\lambda_G}<|q|<\frac{2j\pi}{\lambda_G}$.

\begin{figure}
\includegraphics[width=.99\columnwidth,clip=true]{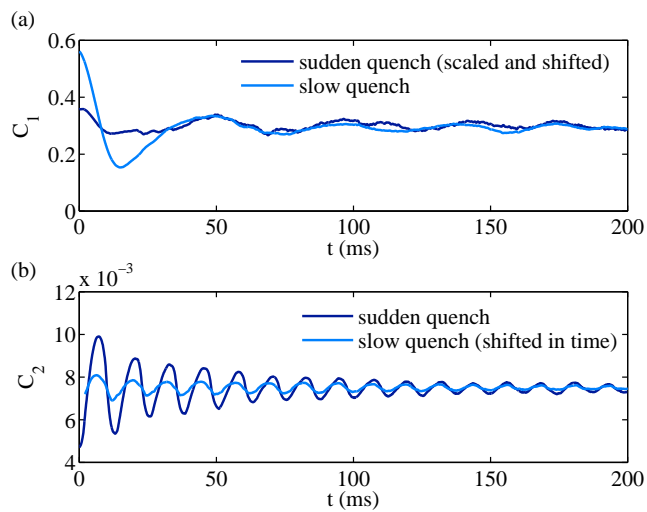}
 \caption{(Color online) Evolution of the quasi-momentum contrast within (a) the first Brillouin zone and (b) the second Brillouin zone after the abrupt and slow quench. For the almost abrupt quench in (a) the contrast ($C_1$) is scaled by a factor of approx.~$-2$ and then shifted by approx.~$-0.6$. For the slow quench in (b) the time is shifted by $2 $ms}
\label{fig:dyn_cont_12_su_sl}
\end{figure}

In Fig.~\ref{fig:dyn_cont_12_su_sl}a the evolution of the contrast within the first Brillouin zone after the almost abrupt quench and slow quench is plotted. The contrast for the almost abrupt quench is scaled by a factor of $\approx -2$ and shifted by $0.6$ms. One sees that the qualitative behaviour of the contrast for both is very similar in particular the oscillation periods are comparable (even though the contrast for the almost abrupt quench is much smaller). We observe damped oscillation with a period of $T \approx 43 $ms. This period is close to the breathing mode of the particles in the trap in the presence of the green lattice. The breathing period corresponds to the new rescaled trap frequency $\Omega\approx\pi\omega_x\sqrt{\frac{J_G}{E_{r,G}}}$ is $T_2 \approx 36.6 $ms. 

In Fig.~\ref{fig:dyn_cont_12_su_sl}b the time evolution of the contrast within the second Brillouin zone of the green lattice after the almost abrupt quench and slow quench is plotted. The time is shifted by $2 $ms for the slow quench. The oscillating periods for two different quenches are very close. Since the particles in the second Brillouin zone are living in the second Bloch band of the green lattice, the tunneling amplitude between neighbouring sites $J_G'$ in the higher band is different. The breathing mode for particles in the second energy band corresponds to the new rescaled trap frequency $\Omega'\approx\pi\omega_x\sqrt{\frac{J_G'}{E_{r,G}}}$. With the tunneling amplitude calculated for the second band ($J_G'=1.9492 \times 10^{-1} E_{r,G}$) the breathing period is $T_3\approx 9.5 $ms that is close to the oscillating period of the contrast in the second Brillouin zone.

\subsection{Discussion of the steady states}
\begin{figure}
\includegraphics[width=.99\columnwidth,clip=true]{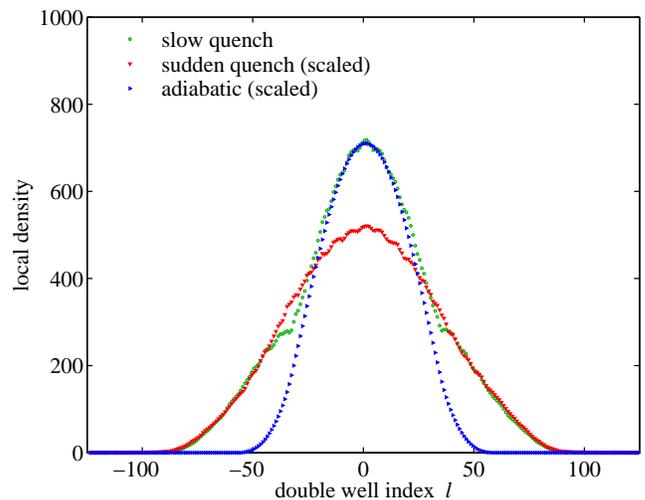}
\caption{(Color online) Density distribution of the stationary state after the slow quench, the almost sudden quench, and the adiabatic quench. The density is integrated over $y$- and $z$-direction and over the wells. }
\label{fig:den_relax_sl_su}
\end{figure}

In this section we investigate the steady states which arise at long times after the abrupt and slow quench. In Fig.~\ref{fig:steady} the energy level occupation after the quenches has already been shown. Since this distribution does not change anymore in the time-independent situation after the quench, this distribution corresponds to the steady state. We had seen that both, the distribution after the sudden and after the abrupt quench had excitations up to relative high energy levels. Further, the distribution after the slow quench has shown a distinct two peak structure of a relatively high peak around low energies and a much broader peak with lower amplitude which extended up to high energies. In this section, we would like to see how this influences local observables as the density distribution. 
The local density of the steady states after the slow and sudden quench are shown in Fig.~\ref{fig:den_relax_sl_su}. Additionally, we display the density distribution after a completely adiabatic quench. 
In both distributions, locally a balanced density has been reached which corresponds to the expected value in thermal equilibrium. However, the global density distribution shows similar features than the energy level occupation. After the slow quench, the global density distribution can be decomposed into two superimposed shapes. In the center the density distribution resembles the stationary state after an adiabatic quench (scaled by $0.73$). Thus, in this region the density seems to come (at least from its form) close to a thermal equilibrium distribution. However, at the boundaries the density distribution resembles the stationary state after the sudden quench (scaled by $0.89$) and its form is very distinct from a thermal distribution. We verified this by comparing the density distribution after the abrupt quench to a thermal distribution varying the temperature. Therefore, the global distribution of the local density signals strongly the non-equilibrium nature of the state.

\section{Experimental detection procedure of the odd-even density contrast}
\label{sec:exp_detect}

\begin{figure}
\includegraphics[width=.99\columnwidth,clip=true]{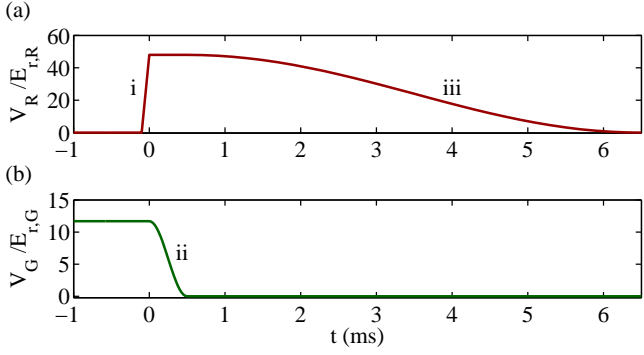}
 \caption{(Color online) Ramp schedule of the depth of a) the red and b) the green lattice potential during the detection procedure of the integrated odd-even contrast. Time zero is chosen to correspond to the beginning of the ramp down of the green lattice.}
\label{fig:exp1_seq}
\end{figure}

\begin{figure}
\includegraphics[width=.99\columnwidth,clip=true]{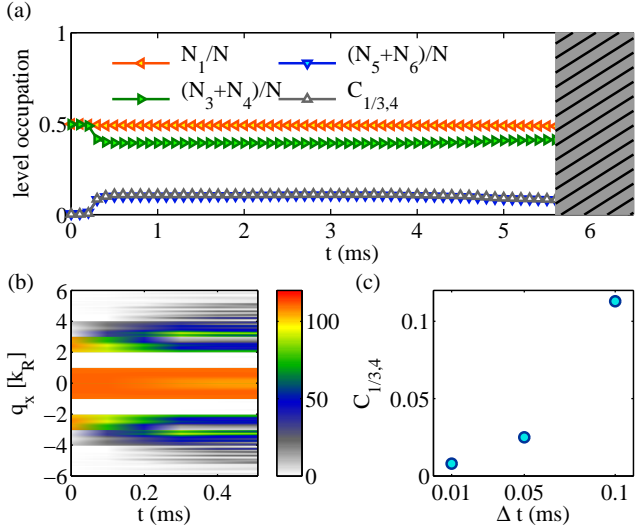}
 \caption{(Color online) a) Time evolution of different fractions of the atoms in different Brillouin zones. The shaded area marks the time regime when some atoms reach the boundary and the simulation does not correspond to experimental results any longer. b) Evolution of the quasi-momentum distribution in $x$-direction integrated over $y$- and $z$-direction for early times of the measurement procedure. The scale of the vertical axis is chosen as in Fig.~\ref{fig:up_trap_px_250dw}. c) The contrast $C_{1/3,4}$ after the green lattice is ramped down during the measurement for different durations of the time steps. The simulations are performed for $N=10^5$ particles in a trap where $450$ double wells are taken into account.}
\label{fig:1st_exp_meas_450dw}
\end{figure}
In order to measure experimentally the initial relaxation discussed in section \ref{sec:shorttime}, the even-odd contrast has been monitored. The experimental detection of the even-odd contrast relies on the use of the bichromatic lattice \cite{Sebbystrabley2007,FoellingBloch2007,GreifEsslinger2013, PertotKollath2014, TrotzkyBloch2012}.
The experimental sequence consists of three parts (see Fig.~\ref{fig:exp1_seq}): i) The red lattice is quickly ($\tau_1=0.1 ms$) ramped up to $V_R=48 E_{r,R}$ (at a phase $\phi=\frac{\pi}{2}$) which produces a superlattice with strong energy offset between neighbouring sites. By this, the occupation of the green lattice well is projected onto the occupation of the corresponding well of the superlattice. 
 ii) The green lattice is ramped down to zero in $\tau_2=0.5 ms$. The parameters for the superlattice are chosen, such that during this ramp down the population of the even wells transfers to the third and fourth Bloch band of the red lattice and the atoms in the odd wells 
remain in the lowest band. The second band remains empty. 
iii) Subsequently an adiabatic band mapping is performed in order to measure the percentage of atoms in the different Bloch bands. This is performed by a slow ramp down of the red lattice in $\tau_3=6 ms$ with an additional time-of-flight imaging of the atoms. 
Thus, the measured time-of-flight images reveal the odd and even lattice well population integrated over the trap. The odd-even contrast ($C_{oe}$) can be measured by the new contrast defined as $C_{1/3,4}=\frac{N1-N3-N4}{N1+N3+N4}$ where $N_j$ is the number of atoms in the $j$-th Brillouin zone.

In order to analyze imperfections of the experimental detection process, we perform numerical simulations starting from the ground state of the green lattice ($V_G=11.7 E_{r,G}, V_R=0,\phi=\frac{\pi}{2}$), i.e.~a balanced density configuration ($C_{oe}=0$).  
The quasi-momentum distribution is calculated at each time step by a direct projection onto the homogeneous quasi-momenta. 

In Fig.~\ref{fig:1st_exp_meas_450dw}a the results of the theoretical simulations of the fraction of atoms in different Brillouin zones and the contrast $C_{1/3,4}$ during the measurement procedure are shown. The shaded area marks when the particles reach the boundary and our simulation is not reliable anymore. For the simulation, a time step of $\Delta t=0.1$ ms is chosen, since this corresponds to the regulation steps of the experiment \cite{PertotKollath2014}. Surprisingly, this discretization time-step causes an increase of the contrast during the ramp down of the green lattice. After the ramp down the contrast remains almost constant. The sharp increase is mainly due to the excitation of atoms from the third and fourth bands to the fifth and sixth bands as can be clearly seen in Fig.~\ref{fig:1st_exp_meas_450dw}b. This increase of the contrast falsifies the detection outcome. In the considered case a contrast of $C_{1/3,4}\approx 0.11$ would be measured instead of the expected zero value of the prepared state. We repeated the simulations for different initial contrasts. The effect of the measurement procedure is strongest on the shown initially balanced situation. On the totally imbalanced situation a discrepancy of about 3\% is found. The slight shift of the zero contrast to a final value has been observed in Ref.~\cite{PertotKollath2014}. This shift could be avoided by decreasing the regulation step of the ramp down of the lattice potential. In order to verify this, we have performed simulations with varying time-step between $\Delta t=0.1$ ms to $0.01$ ms and we find that the fraction of atoms excited to the higher bands decreases to $C_{1/3,4}\approx 0.01$ (see Fig.~\ref{fig:1st_exp_meas_450dw}c).

\section{Conclusion}

In this work we have studied theoretically the loading procedure of non-interacting fermions into a one-dimensional superlattice, the relaxation after an abrupt change of the superlattice, and the measurement procedure of the odd-even density contrast. Our calculations have taken into account the full three dimensional and spatially inhomogeneous nature of the cloud subjected to an optical superlattice potential at the expense of considering non-interacting fermions. The findings we have obtained led to insights which can be used to improve the experimental loading and detection procedures. Concerning the loading ramp of the fermions into the optical lattice, in our results a large mixing and the creation of excitations has been observed only in the very initial time of the loading procedure until the Bloch bands become separated (even in the presence of a trap). Thus, the initial loading has to be optimized in an experiment, whereas after the opening of the gap a faster ramp can be applied without harm. Additionally for fermionic atoms --in contrast to bosonic atoms-- the number of atoms and the trap configurations plays an important role, since due to the Pauli blocking each available energy level can only be occupied by one fermion (or two for the spinful situation). Starting from a certain density the higher bands have to contribute and a loading of these cannot be avoided even by a locally adiabatic loading procedure. The solution to this problem is either to start with a low density or enough available levels in the directions perpendicular to the lattice direction. 
The trap has the additional problem that due to the lattice ramp up a change of the global density distribution is induced. Another way of seeing the strong influence of the trap is the induced level crossing during the ramp. Controlling the frequency of the trap this effect can be reduced. The redistribution of the density takes typically a long time. The redistribution at the boundaries can be almost blocked by Bloch oscillations due to the local energy offset. In the considered case of non-interacting fermions this redistribution is even at infinite time prohibited by the conservation laws and it will never relax to its thermal equilibrium value. Even though, this is an artefact of the integrable situation, some of these integrable features might also dominate the situation in which the integrability is slightly broken by a weak interaction \footnote{Such a reminiscient behaviour to the integrable system has for example been observed for interaction quenches in bosonic \cite{KollathAltman2007} and fermionic systems \cite{MoeckelKehrein2008, MoeckelKehrein2010, KollarWerner2009} .}. Let us mention that a very slow redistribution had already been identified previously, for the interacting case. In the interacting model Mott-insulating barriers can arise which can almost totally block the redistribution of the atoms \cite{BernierKollath2011,Natu2011}. Thus in experiments this is a crucial point which has to be considered. A corresponding adjustment of the trapping potential, as it is partially the case in red-detuned lattices \footnote{For red-detuned lattice potentials since an additional trapping potential induced by the focusing of the lattice beams decreases the expected redistribution.}, could cure this problem (cf.~also recent quasi-exact simulations for bosonic atoms in one-dimension which support this suggestion \cite{DolfiTroyer2014}).  

The relaxation dynamics after a sudden and slow change of the bichromatic potential to a monochromatic potential has shown that locally a fast relaxation can occur for some observables, whereas the global relaxation takes place on a much longer time scale. Due to the integrability of our system, the long-time steady state could show a structure of two superimposed peaks with different width, which clearly does not correspond to a thermal equilibrium state. 
 We identified different important time-scales. The local relaxation has taken place on the time-scale of the tunneling time, here given by $\tau_1=\frac{2\pi \hbar}{4J_G} \approx 1.08 $ms. Additionally, the global density redistribution occurs on a longer time scale. An impression of these give the breathing modes, which are approximately one to two orders of magnitudes larger. 

In the detection procedure for the odd-even density contrast also the time-scales play a very important role. We have found that a fast switch off of the green lattice potential and dominantly a relatively long regulation step of the lattice potential influences the resulting contrast. Whereas this change is negligible for a high contrast, in the situation of vanishing contrast the results are falsified and a finite contrast is found. This can be cured in the experimental results by a change of the lattice ramp time and the regulation step.  

\appendix
\section{Comparison of continuous simulations versus discrete simulations}
\label{app:tightbinding}
In this appendix we describe the simulations performed in the discrete tight-binding approximation shown in Fig.~\ref{fig:breathing}. For the strongly imbalanced superlattice along $x$-direction, the Hamiltonian of the system is 
\begin{eqnarray}
H_x&=&\sum_{l=-L/2}^{L/2}[-J_{12} c_{2l-1}^\dagger c_{2l}- J_{23}c_{2l}^\dagger c_{2l+1}\nonumber\\
&-&J_{13}c_{2l-1}^\dagger c_{2l+1}-J_{24}c_{2l}^\dagger c_{2l+2} +h.c.]\nonumber\\
&+& \sum_{j=-L}^{L} \left[\frac{\Delta}{2} (-1)^j c_{j}^\dagger c_{j} + V_t j^2 c_{j}^\dagger c_{j}\right].
\label{eq:Hx_dis_before_quench}
\end{eqnarray}
Here we label each double well with the index $l$ and the index for the wells inside the double well $l$ are $j=2l-1$ and $j=2l$. We restricted the description to include hopping terms up to the second nearest neighbours. Here $J_{m n}$ is the hopping amplitude from the well $j=m$ to the well $j=n$. $V_t$ is the trap potential in the discrete form and $\Delta$ is the energy offset between neighbouring sites in one double well without the trap.
With the parameters we used for the continuous simulation the energy offset in the initially prepared state is given by $\Delta=15 E_{r,R}$ which is much larger than the tunneling amplitudes. In the simulations we calculate the ground state of the Hamiltonian where $H_x$ is as Eq.~\ref{eq:Hx_dis_before_quench} and $H_y$ and $H_z$ as in Eq.~\ref{eq:Hxyz_time_sec_quant} with the corresponding trap frequency when both of the red and green lattice are on.

After the quench when only the green lattice is on and there is no double well structure any more, the Hamiltonian in $x$-direction is given by
\begin{eqnarray*}
H_x&=&\sum_{j=-L}^{L}[-J_{G} c_{j}^\dagger c_{j+1}-J_{G} c_{j+1}^\dagger c_{j}+V_t j^2 c_{j}^\dagger c_{j}].
\end{eqnarray*}
Here $J_G=1.3085\times 10^{-2}E_{r,G}$ is the hopping in the remaining green lattice. The trap potential $V_t$ is the same as before the quench. $H_y$ and $H_z$ are the same as given in Eq. \ref{eq:Hxyz_time_sec_quant} with the corresponding trap frequency for the green lattice. 

\begin{figure}
\includegraphics[width=.99\columnwidth,clip=true]{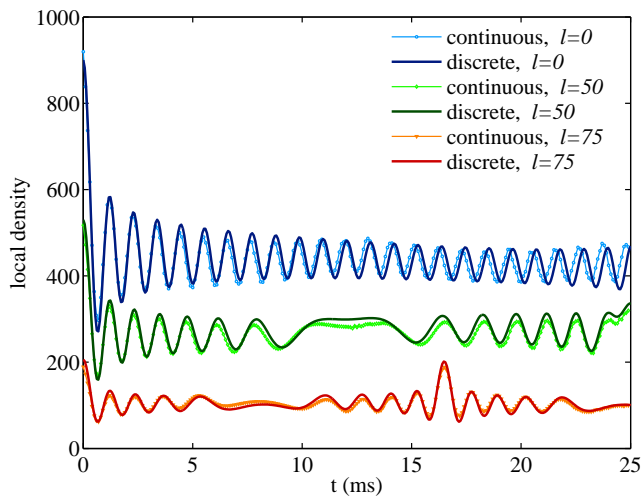}
 \caption{(Color online) Evolution of the local density of odd wells at different regions of the trap after the abrupt quench compared for the tight-binding and continuous model. It is simulated for $10^5$ particles in the trap with $250$ double wells.}
\label{fig:dis_con_dens_250dw_diffx}
\end{figure}

In Fig.~\ref{fig:dis_con_dens_250dw_diffx} we compare the local density evolution of the continuous (Eq.~\ref{eq:Hxyz_time_sec_quant}) and tight-binding model (Eq.~\ref{eq:Hx_dis_before_quench}) starting from the ground state of the imbalanced bichromatic lattice before the quench. The evolution of the local density is plotted for different wells in the trap. The density is summed over the $y$- and $z$-direction and for the continuous case the density is also summed over each well. One can see that the behaviour for the discrete and the continuous model are very similar only slight differences in the frequency are seen which start to cause deviations after a time of $5 $ms. This we attribute to either the discretization of the continuous model or the approximate hoppings ($J_{mn}$) calculated from the band structure. 

\begin{thebibliography}{10}

\bibitem{KetterleZwierlein2008}
M.~W.~Z. Wolfgang~Ketterle,
\newblock Proceedings of the International School of Physics "Enrico Fermi",
  Course CLXIV, Varenna, 20 - 30 June 2006, edited by M. Inguscio, W. Ketterle,
  and C. Salomon (IOS Press, Amsterdam) .

\bibitem{BlochZwerger2008}
I.~Bloch, J.~Dalibard, and W.~Zwerger,
\newblock Rev. Mod. Phys. {\bf 80}, 885 (2008).

\bibitem{Schneider2008}
U.~Schneider {\em et~al.},
\newblock Science {\bf 322}, 1520 (2008).

\bibitem{Joerdens2008}
R.~J\"ordens, N.~Strohmaier, K.~Gunter, H.~Moritz, and T.~Esslinger,
\newblock Nature {\bf 455}, 204 (2008).

\bibitem{Lewenstein2007}
M.~Lewenstein {\em et~al.},
\newblock Advances in Physics {\bf 56}, 243 (2007).

\bibitem{JoerdensTroyer2010}
R.~J\"ordens {\em et~al.},
\newblock Phys. Rev. Lett. {\bf 104}, 180401 (2010).

\bibitem{DeLeoParcollet2008}
L.~De~Leo, C.~Kollath, A.~Georges, M.~Ferrero, and O.~Parcollet,
\newblock Phys. Rev. Lett. {\bf 101}, 210403 (2008).

\bibitem{PolletTroyer2008}
L.~Pollet, C.~Kollath, K.~V. Houcke, and M.~Troyer,
\newblock New Journal of Physics {\bf 10}, 065001 (2008).

\bibitem{DolfiTroyer2014}
{Michele Dolfi, Adrian Kantian, Bela Bauer, Matthias Troyer},
\newblock cond-mat/1410.5829.

\bibitem{PolkovnikovSilva2011}
A.~Polkovnikov, K.~Sengupta, A.~Silva, and M.~Vengalattore,
\newblock Rev. Mod. Phys. {\bf 83}, 863 (2011).

\bibitem{Igloi2000}
F.~Igl\'oi and H.~Rieger,
\newblock Phys. Rev. Lett. {\bf 85}, 3233 (2000).

\bibitem{CalabreseCardy2006}
P.~Calabrese and J.~Cardy,
\newblock Phys.~ Rev.~ Lett. {\bf 96}, 136801 (2006).

\bibitem{Cazalilla2006}
M.~A. Cazalilla,
\newblock Phys. Rev. Lett. {\bf 97}, 156403 (2006).

\bibitem{Rigol2006}
M.~Rigol, A.~Muramatsu, and M.~Olshanii,
\newblock Phys. Rev. A {\bf 74}, 053616 (2006).

\bibitem{KollathAltman2007}
C.~Kollath, A.~M. L\"auchli, and E.~Altman,
\newblock Phys. Rev. Lett. {\bf 98}, 180601 (2007).

\bibitem{ManmanaMuramatsu2007}
S.~R. Manmana, S.~Wessel, R.~M. Noack, and A.~Muramatsu,
\newblock Phys. Rev. Lett. {\bf 98}, 210405 (2007).

\bibitem{BarmettlerRey2008}
P.~Barmettler {\em et~al.},
\newblock Phys. Rev. A {\bf 78}, 012330 (2008).

\bibitem{RigolOlshani2008}
M.~Rigol, V.~Dunjko, and M.~Olshanii,
\newblock Nature {\bf 452}, 854 (2008).

\bibitem{Lauchli2008}
A.~M. L\"auchli and C.~Kollath,
\newblock Journal of Statistical Mechanics: Theory and Experiment {\bf 2008},
  P05018 (2008).

\bibitem{Cubitt2008}
T.~S. Cubitt and J.~I. Cirac,
\newblock Phys. Rev. Lett. {\bf 100}, 180406 (2008).

\bibitem{BarthelSchollwoeck2008}
T.~Barthel and U.~Schollw\"ock,
\newblock Phys. Rev. Lett. {\bf 100}, 100601 (2008).

\bibitem{Rossini2009}
D.~Rossini, A.~Silva, G.~Mussardo, and G.~E. Santoro,
\newblock Phys. Rev. Lett. {\bf 102}, 127204 (2009).

\bibitem{Manmana2009}
S.~R. Manmana, S.~Wessel, R.~M. Noack, and A.~Muramatsu,
\newblock Phys. Rev. B {\bf 79}, 155104 (2009).

\bibitem{IucciCazalilla2009}
A.~Iucci and M.~A. Cazalilla,
\newblock Phys. Rev. A {\bf 80}, 063619 (2009).

\bibitem{MoeckelKehrein2009}
M.~Moeckel and S.~Kehrein,
\newblock Annals of Physics {\bf 324}, 2146  (2009).

\bibitem{Uhrig2009}
G.~S. Uhrig,
\newblock Phys. Rev. A {\bf 80}, 061602 (2009).

\bibitem{SotiriadisCardy2010}
S.~Sotiriadis and J.~Cardy,
\newblock Phys. Rev. B {\bf 81}, 134305 (2010).

\bibitem{Mathey2010}
L.~Mathey and A.~Polkovnikov,
\newblock Phys. Rev. A {\bf 81}, 033605 (2010).

\bibitem{BarmettlerAltman2010}
P.~Barmettler, M.~Punk, V.~Gritsev, E.~Demler, and E.~Altman,
\newblock New Journal of Physics {\bf 12}, 055017 (2010).

\bibitem{KollarEckstein2011}
M.~Kollar, F.~A. Wolf, and M.~Eckstein,
\newblock Phys. Rev. B {\bf 84}, 054304 (2011).

\bibitem{Enss2012}
T.~Enss and J.~Sirker,
\newblock New Journal of Physics {\bf 14}, 023008 (2012).

\bibitem{Goth2012}
F.~Goth and F.~F. Assaad,
\newblock Phys. Rev. B {\bf 85}, 085129 (2012).

\bibitem{MosselCaux2012}
J.~Mossel and J.-S. Caux,
\newblock New Journal of Physics {\bf 14}, 075006 (2012).

\bibitem{BarmettlerKollath2012}
P.~Barmettler, D.~Poletti, M.~Cheneau, and C.~Kollath,
\newblock Phys. Rev. A {\bf 85}, 053625 (2012).

\bibitem{Natu2013}
S.~S. Natu and E.~J. Mueller,
\newblock Phys. Rev. A {\bf 87}, 063616 (2013).

\bibitem{EsslerKehrein2014}
F.~H.~L. Essler, S.~Kehrein, S.~R. Manmana, and N.~J. Robinson,
\newblock Phys. Rev. B {\bf 89}, 165104 (2014).

\bibitem{Queisser2014}
F.~Queisser, K.~V. Krutitsky, P.~Navez, and R.~Sch\"utzhold,
\newblock Phys. Rev. A {\bf 89}, 033616 (2014).

\bibitem{AokiWerner2014}
H.~Aoki {\em et~al.},
\newblock Rev. Mod. Phys. {\bf 86}, 779 (2014).

\bibitem{GreinerBloch2002b}
M.~Greiner, O.~Mandel, T.~W. Hansch, and I.~Bloch,
\newblock Nature {\bf 419}, 51 (2002).

\bibitem{Sebbystrabley2007}
J.~Sebby-Strabley {\em et~al.},
\newblock Phys. Rev. Lett. {\bf 98}, 200405 (2007).

\bibitem{FoellingBloch2007}
S.~F\"olling {\em et~al.},
\newblock Nature {\bf 448}, 1029 (2007).

\bibitem{BakrGreiner2009}
W.~S. Bakr, J.~I. Gillen, A.~Peng, S.~F\"olling, and M.~Greiner,
\newblock Nature {\bf 462}, 74 (2009).

\bibitem{HungChin2010}
C.-L. Hung, X.~Zhang, N.~Gemelke, and C.~Chin,
\newblock Phys. Rev. Lett. {\bf 104}, 160403 (2010).

\bibitem{ShersonKuhr2010}
J.~F. Sherson {\em et~al.},
\newblock Nature {\bf 467}, 68 (2010).

\bibitem{ChenDeMarco2011}
D.~Chen, M.~White, C.~Borries, and B.~DeMarco,
\newblock Phys. Rev. Lett. {\bf 106}, 235304 (2011).

\bibitem{RonzheimerBloch2013}
J.~P. Ronzheimer {\em et~al.},
\newblock Phys. Rev. Lett. {\bf 110}, 205301 (2013).

\bibitem{KinoshitaWeiss2006}
T.~Kinoshita, T.~Wenger, and D.~S. Weiss,
\newblock Nature {\bf 440}, 900 (2006).

\bibitem{GringSchmiedmayer2012}
M.~Gring {\em et~al.},
\newblock Science {\bf 337}, 1318 (2012).

\bibitem{Cheneau2012}
M.~Cheneau {\em et~al.},
\newblock Nature {\bf 481}, 484 (2012).

\bibitem{TrotzkyBloch2012}
S.~Trotzky {\em et~al.},
\newblock Nat. Phys. {\bf 8}, 325 (2012).

\bibitem{CramerEisert2008}
M.~Cramer, A.~Flesch, I.~P. McCulloch, U.~Schollw\"ock, and J.~Eisert,
\newblock Phys. Rev. Lett. {\bf 101}, 063001 (2008).

\bibitem{FleschEisert2008}
A.~Flesch, M.~Cramer, I.~P. McCulloch, U.~Schollw\"ock, and J.~Eisert,
\newblock Phys. Rev. A {\bf 78}, 033608 (2008).

\bibitem{Strohmaier2007}
N.~Strohmaier {\em et~al.},
\newblock Phys. Rev. Lett. {\bf 99}, 220601 (2007).

\bibitem{Hackermueller2010}
L.~Hackerm\"uller {\em et~al.},
\newblock Science {\bf 327}, 1621 (2010).

\bibitem{Schneider2012}
U.~Schneider {\em et~al.},
\newblock Nat. Phys. {\bf 8}, 213 (2012).

\bibitem{PertotKollath2014}
D.~Pertot {\em et~al.},
\newblock Phys. Rev. Lett. {\bf 113}, 170403 (2014).

\bibitem{KrauserSengstock2012}
J.~S. Krauser {\em et~al.},
\newblock Nat. Phys. {\bf 8}, 813 (2012).

\bibitem{KoschorreckKoehl2013}
M.~Koschorreck, D.~Pertot, E.~Vogt, and M.~K\"ohl,
\newblock Nat. Phys. {\bf 9}, 405 (2013),
\newblock Letter.

\bibitem{GuenterEsslinger2005}
K.~G\"unter, T.~St\"oferle, H.~Moritz, M.~K\"ohl, and T.~Esslinger,
\newblock Phys. Rev. Lett. {\bf 95}, 230401 (2005).

\bibitem{FroehlichKoehl2011}
B.~Fr\"ohlich {\em et~al.},
\newblock Phys. Rev. Lett. {\bf 106}, 105301 (2011).

\bibitem{Martiyanov2010}
K.~Martiyanov, V.~Makhalov, and A.~Turlapov,
\newblock Phys. Rev. Lett. {\bf 105}, 030404 (2010).

\bibitem{SommerZwierlein2012}
A.~T. Sommer, L.~W. Cheuk, M.~J.~H. Ku, W.~S. Bakr, and M.~W. Zwierlein,
\newblock Phys. Rev. Lett. {\bf 108}, 045302 (2012).

\bibitem{GreifEsslinger2013}
D.~Greif, T.~Uehlinger, G.~Jotzu, L.~Tarruell, and T.~Esslinger,
\newblock Science {\bf 340}, 1307 (2013).

\bibitem{Note1}
Such a reminiscient behaviour to the integrable system has for example been
  observed for interaction quenches in bosonic \cite {KollathAltman2007} and
  fermionic systems \cite {MoeckelKehrein2008, MoeckelKehrein2010,
  KollarWerner2009} .

\bibitem{BernierKollath2011}
J.-S. Bernier, G.~Roux, and C.~Kollath,
\newblock Phys. Rev. Lett. {\bf 106}, 200601 (2011).

\bibitem{Natu2011}
S.~S. Natu, K.~R.~A. Hazzard, and E.~J. Mueller,
\newblock Phys. Rev. Lett. {\bf 106}, 125301 (2011).

\bibitem{Note2}
For red-detuned lattice potentials since an additional trapping potential
  induced by the focusing of the lattice beams decreases the expected
  redistribution.

\bibitem{MoeckelKehrein2008}
M.~Moeckel and S.~Kehrein,
\newblock Phys. Rev. Lett. {\bf 100}, 175702 (2008).

\bibitem{MoeckelKehrein2010}
M.~Moeckel and S.~Kehrein,
\newblock New Journal of Physics {\bf 12}, 055016 (2010).

\bibitem{KollarWerner2009}
M.~Eckstein, M.~Kollar, and P.~Werner,
\newblock Phys. Rev. Lett. {\bf 103}, 056403 (2009).

\end{thebibliography}
\end{document}